\PassOptionsToPackage{unicode}{hyperref}
\PassOptionsToPackage{hyphens}{url}
\PassOptionsToPackage{dvipsnames,svgnames,x11names}{xcolor}
\documentclass[
]{article}
\usepackage{amsmath,amssymb}
\usepackage{lmodern}
\usepackage{iftex}
\ifPDFTeX
  \usepackage[T1]{fontenc}
  \usepackage[utf8]{inputenc}
  \usepackage{textcomp} 
\else 
  \usepackage{unicode-math}
  \defaultfontfeatures{Scale=MatchLowercase}
  \defaultfontfeatures[\rmfamily]{Ligatures=TeX,Scale=1}
\fi
\IfFileExists{upquote.sty}{\usepackage{upquote}}{}
\IfFileExists{microtype.sty}{
  \usepackage[]{microtype}
  \UseMicrotypeSet[protrusion]{basicmath} 
}{}
\makeatletter
\@ifundefined{KOMAClassName}{
  \IfFileExists{parskip.sty}{%
    \usepackage{parskip}
  }{
    \setlength{\parindent}{0pt}
    \setlength{\parskip}{6pt plus 2pt minus 1pt}}
}{
  \KOMAoptions{parskip=half}}
\makeatother
\usepackage{xcolor}
\usepackage[margin=1in]{geometry}
\usepackage{color}
\usepackage{fancyvrb}

\DefineVerbatimEnvironment{Highlighting}{Verbatim}{commandchars=\\\{\}}
\usepackage{framed}
\definecolor{shadecolor}{RGB}{248,248,248}
\newenvironment{Shaded}{\begin{snugshade}}{\end{snugshade}}

\newcommand{\AttributeTok}[1]{\textcolor[rgb]{0.77,0.63,0.00}{#1}}

\newcommand{\ConstantTok}[1]{\textcolor[rgb]{0.00,0.00,0.00}{#1}}
\newcommand{\ControlFlowTok}[1]{\textcolor[rgb]{0.13,0.29,0.53}{\textbf{#1}}}

\newcommand{\DecValTok}[1]{\textcolor[rgb]{0.00,0.00,0.81}{#1}}

\newcommand{\FloatTok}[1]{\textcolor[rgb]{0.00,0.00,0.81}{#1}}
\newcommand{\FunctionTok}[1]{\textcolor[rgb]{0.00,0.00,0.00}{#1}}

\newcommand{\NormalTok}[1]{#1}

\newcommand{\OtherTok}[1]{\textcolor[rgb]{0.56,0.35,0.01}{#1}}

\newcommand{\SpecialCharTok}[1]{\textcolor[rgb]{0.00,0.00,0.00}{#1}}

\usepackage{longtable,booktabs,array}
\usepackage{calc} 
\usepackage{etoolbox}
\makeatletter
\patchcmd\longtable{\par}{\if@noskipsec\mbox{}\fi\par}{}{}
\makeatother
\IfFileExists{footnotehyper.sty}{\usepackage{footnotehyper}}{\usepackage{footnote}}
\makesavenoteenv{longtable}
\usepackage{graphicx}
\makeatletter
\def\maxwidth{\ifdim\Gin@nat@width>\linewidth\linewidth\else\Gin@nat@width\fi}
\def\maxheight{\ifdim\Gin@nat@height>\textheight\textheight\else\Gin@nat@height\fi}
\makeatother
\setkeys{Gin}{width=\maxwidth,height=\maxheight,keepaspectratio}
\makeatletter
\def\fps@figure{htbp}
\makeatother
\ifLuaTeX
\usepackage[bidi=basic]{babel}
\else
\usepackage[bidi=default]{babel}
\fi
\babelprovide[main,import]{american}

\def\languageshorthands#1{}
\usepackage{hyperref}
\definecolor{linkcolor}{HTML}{D55E00}
\definecolor{citecolor}{HTML}{009E73}
\definecolor{urlcolor}{HTML}{0072B2}
\hypersetup{
    colorlinks,
    linkcolor={linkcolor},
    citecolor={citecolor},
    urlcolor={urlcolor}
}

\makeatletter
\@ifundefined{DecValTok}{}{
  \renewcommand{\DecValTok}[1]{\textcolor[HTML]{009E73}{#1}}
  \renewcommand{\FloatTok}[1]{\textcolor[HTML]{009E73}{#1}}
  \renewcommand{\ConstantTok}[1]{\textcolor[HTML]{009E73}{#1}}
  \renewcommand{\ControlFlowTok}[1]{\textcolor[HTML]{0072B2}{\textbf{#1}}}
  \renewcommand{\OtherTok}[1]{\textcolor[HTML]{000000}{#1}}
  
  \renewcommand{\AttributeTok}[1]{\textcolor[HTML]{CC79A7}{#1}}
  \renewcommand{\FunctionTok}[1]{\textcolor[HTML]{56B4E9}{#1}}
}
\makeatother

\usepackage{caption}
\usepackage{enumitem}
\setlist[itemize]{topsep=0pt}

\usepackage[vlined]{algorithm2e}
\usepackage{algorithmic}

\usepackage[absolute]{textpos}
\newcommand{\BigO}{\mathcal{O}}
\newcommand{\SD}{\operatorname{SD}}
\newcommand{\Sn}{\operatorname{S}_n}
\newcommand{\Qn}{\operatorname{Q}_n}
\newcommand{\MAD}{\operatorname{MAD}}
\newcommand{\med}{\operatorname{median}}
\newcommand{\lomed}{\operatorname{lomedian}}
\newcommand{\himed}{\operatorname{himedian}}
\newcommand{\E}{\mathbb{E}}
\newcommand{\V}{\mathbb{V}}
\usepackage{amsmath}
\usepackage{float}
\usepackage{csquotes}
\usepackage{booktabs}
\usepackage{longtable}
\usepackage{array}
\usepackage{multirow}
\usepackage{wrapfig}
\usepackage{colortbl}
\usepackage{pdflscape}
\usepackage{tabu}
\usepackage{threeparttable}
\usepackage{threeparttablex}
\usepackage[normalem]{ulem}
\usepackage{makecell}
\usepackage{xcolor}
\ifLuaTeX
  \usepackage{selnolig}  
\fi
\usepackage[style=alphabetic,sorting=anyt]{biblatex}
\IfFileExists{bookmark.sty}{\usepackage{bookmark}}{\usepackage{hyperref}}
\IfFileExists{xurl.sty}{\usepackage{xurl}}{} 
\hypersetup{
  pdftitle={Finite-sample Rousseeuw-Croux scale estimators},
  pdfauthor={Andrey Akinshin},
  pdflang={en-US},
  colorlinks=true,
  linkcolor={linkcolor},
  filecolor={Maroon},
  citecolor={citecolor},
  urlcolor={urlcolor},
  pdfcreator={LaTeX via pandoc}}

\title{Finite-sample Rousseeuw-Croux scale estimators}
\author{Andrey Akinshin\\
Huawei Research, \href{mailto:andrey.akinshin@gmail.com}{\nolinkurl{andrey.akinshin@gmail.com}}}
\date{}

\begin{document}
\maketitle
\begin{abstract}
The Rousseeuw-Croux \(S_n\), \(Q_n\) scale estimators and the median absolute deviation \(\MAD_n\)
can be used as consistent estimators for the standard deviation under normality.
All of them are highly robust: the breakdown point of all three estimators is \(50\%\).
However, \(S_n\) and \(Q_n\) are much more efficient than~\(\MAD_n\):
their asymptotic Gaussian efficiency values are \(58\%\) and \(82\%\) respectively compared to \(37\%\) for~\(\MAD_n\).
Although these values look impressive, they are only asymptotic values.
The actual Gaussian efficiency of \(S_n\) and \(Q_n\) for small sample sizes
is noticeable lower than in the asymptotic case.

The original work by Rousseeuw and Croux (1993)
provides only rough approximations of the finite-sample bias-correction factors for \(S_n\), \(Q_n\)
and brief notes on their finite-sample efficiency values.
In this paper, we perform extensive Monte-Carlo simulations in order to obtain refined values of the
finite-sample properties of the Rousseeuw-Croux scale estimators.
We present accurate values of the bias-correction factors and Gaussian efficiency for small samples (\(n \leq 100\))
and prediction equations for samples of larger sizes.

\end{abstract}


\hypertarget{sec:intro}{%
\section{Introduction}\label{sec:intro}}

The standard deviation is a classic measure of statistical dispersion which is used to describe the normal distribution.
Unfortunately, it is not robust: a single extreme outlier can completely distort the standard deviation estimations.
That is why the robust statistics provide a plethora of robust dispersion estimators
which are resistant to gross errors.
With proper scale constants,
these estimators can be asymptotically consistent estimators for the standard deviation under normality.

One of the most popular robust dispersion estimators is the median absolute deviation around the median.
For a sample \(x = \{ x_1, x_2, \ldots, x_n \}\) of i.i.d. random variables, it is defined as follows:

\[
\MAD_n(x) = B_n \cdot \med_i |x_i - \med_j x_j|,
\]

where \(\med\) is the sample median, \(B_n\) is a scale constant.
We use \(B_n\) to make \(\MAD_n\) an unbiased Fisher-consistent (see \autocite{fisher1922}) estimator
for the standard deviation under the normal distribution.
The asymptotic value of \(B_n\) is well-known and given by

\[
B_\infty = \frac{1}{\Phi^{-1}(3/4)} \approx 1.4826022185056 \approx 1.4826,
\]

where \(\Phi^{-1}\) is the quantile function of the standard normal distribution.
For finite samples, we have to introduce bias-correction factors \(b_n\)
(approximated values can be found in \autocite{akinshin2022madfactors} and \autocite{park2020})
such that \(B_n = B_\infty \cdot b_n\).

The median absolute deviation is considered as a robust replacement for the standard deviation
in various statistical textbooks including
\autocite{mosteller1977,hampel1986,huber2009,wilcox2016,maronna2019,jureckova2019}.
It has the highest possible breakdown point of \(50\%\) (see \autocite[p14]{rousseeuw1987}).
However, its asymptotic relative statistical efficiency to the standard deviation under normality (Gaussian efficiency)
is only \(36.75\%\).
Thus, a straightforward replacement of the standard deviation by the median absolute deviation
significantly reduces the statistical efficiency of the used dispersion estimator.
This leads to a poor precision level of the obtained estimations.

There are numerous ways to increase efficiency.
For example, we can estimate the \(\MAD_n\) using more efficient median estimators
like the Harrell-Davis quantile estimator (see \autocite{harrell1982}) or its trimmed modification (see \autocite{akinshin2022thdqe}).
Using adjusted scale constants (see \autocite{akinshin2022madfactors}),
the updated \(\MAD_n\) can be also used as a consistent estimator for the standard deviation under normality.
This approach has slightly higher Gaussian efficiency and lower breakdown point.
It is reasonable to use it only if we are interested in the accurate estimations of the \(\MAD\) itself.
However, it is not the optimal way to estimate the standard deviation.
There are also other alternatives like
the quantile absolute deviation (\autocite{akinshin2022qad}),
the interquartile and interdecile ranges,
the trimmed standard deviation (see \autocite{lax1985}),
the Winsorized standard deviation (see \autocite{wilcox2016}),
the biweight midvariance (see \autocite{lax1985}),
the Shamos estimator (see \autocite{shamos1976}),
and others (e.g., see \autocite{daniell1920}).
Each of these estimators maintains its own balance between statistical efficiency and robustness.
Typically, higher efficiency is achieved by lowering the value of the breakdown point.

In \autocite{rousseeuw1993}, Peter J. Rousseeuw and Christophe Croux suggested using two new scale estimators: \(S_n\) and~\(Q_n\).
Both of them have the breakdown point of \(50\%\), but they are more efficient than \(\MAD_n\):
the asymptotic Gaussian efficiency of \(S_n\) and \(Q_n\) are \(58\%\) and \(82\%\) respectively.
Such exceptional efficiency makes them decent replacements
for the median absolute deviation as robust measures of scale.
While the algorithmic complexity of straightforward implementations for \(S_n\) and \(Q_n\) is \(\BigO(n^2)\),
\autocite{croux1992} presents a fast evaluation algorithm that takes only \(\BigO(n \log n)\).
A fast algorithm for online computation of \(Q_n\) can be found in \autocite{cafaro2020}.
In \autocite{smirnov2014},
a parametric family of M-estimators of scale based on the \(Q_n\) estimator is presented,
which allows obtaining even higher statistical efficiency.

The \(S_n\) estimator is given by

\[
S_n(x) = C_n \cdot \lomed_i \; \himed_j \; |x_i - x_j|,
\]

where
\(\lomed\) and \(\himed\) are the \(\lfloor (n+1) / 2 \rfloor^\textrm{th}\) and
\((\lfloor n / 2 \rfloor + 1)^\textrm{th}\) order statistics out of \(n\) numbers,
\(C_n\) is a scale constant that we use
to make \(S_n\) an unbiased estimator for the standard deviation under normality.
The asymptotic value of \(C_n\) can be obtained (see \autocite[p.1275, Theorem 2]{rousseeuw1993}) as a solution of the equation
\(\Phi(\Phi^{-1}(3/4) + 1 / C_\infty) - \Phi(\Phi^{-1}(3/4) - 1 / C_\infty) = 1/2\).
Its approximated value is

\[
C_\infty \approx 1.19259855312321 \approx 1.1926.
\]

The \(Q_n\) estimator is given by

\[
Q_n(x) = D_n \cdot \{ |x_i-x_j|;\; i < j \}_{(k)},
\]

where
\(\{\cdot\}_{(k)}\) is the \(k^\textrm{th}\) order statistic of the given set,
\(k = \binom{\lfloor n / 2 \rfloor + 1}{2}\) (asymptotically, it converges to the first quartile),
\(D_n\) is a scale constant that we use
to make \(Q_n\) an unbiased estimator for the standard deviation under normality.
The asymptotic value of \(D_n\) is given by (see \autocite[p.1277]{rousseeuw1993})

\[
D_\infty = \frac{1}{\sqrt{2} \Phi^{-1}(5/8)} \approx 2.21914446598508 \approx 2.2191.
\]

Note that \autocite{croux1992}, \autocite{rousseeuw1993}, and some other papers contain a typo:
they use \(D_\infty = 2.2219\) instead of 2.2191.
This slight mistake is widespread across various papers and statistical packages.
In this paper, we use the correct value of \(2.2191\).

\bigskip
\clearpage

For finite samples, we have to introduce bias-correction factors \(c_n\) and \(d_n\)
such that \(C_n = C_\infty \cdot c_n\), \(D_n = D_\infty \cdot d_n\).
In \autocite{croux1992}, rough approximations of \(c_n\) and \(d_n\) are given.
The values for \(n \leq 9\) from \autocite{croux1992} are presented in Table~\ref{tab:factors-rc}.

\begin{longtable}[]{@{}
  >{\raggedright\arraybackslash}p{(\columnwidth - 16\tabcolsep) * \real{0.1111}}
  >{\raggedright\arraybackslash}p{(\columnwidth - 16\tabcolsep) * \real{0.1111}}
  >{\raggedright\arraybackslash}p{(\columnwidth - 16\tabcolsep) * \real{0.1111}}
  >{\raggedright\arraybackslash}p{(\columnwidth - 16\tabcolsep) * \real{0.1111}}
  >{\raggedright\arraybackslash}p{(\columnwidth - 16\tabcolsep) * \real{0.1111}}
  >{\raggedright\arraybackslash}p{(\columnwidth - 16\tabcolsep) * \real{0.1111}}
  >{\raggedright\arraybackslash}p{(\columnwidth - 16\tabcolsep) * \real{0.1111}}
  >{\raggedright\arraybackslash}p{(\columnwidth - 16\tabcolsep) * \real{0.1111}}
  >{\raggedright\arraybackslash}p{(\columnwidth - 16\tabcolsep) * \real{0.1111}}@{}}
\caption{\label{tab:factors-rc} \(S_n\) and \(Q_n\) finite-sample bias-correction factors from \autocite{croux1992}.}\tabularnewline
\toprule()
\begin{minipage}[b]{\linewidth}\raggedright
n
\end{minipage} & \begin{minipage}[b]{\linewidth}\raggedright
2
\end{minipage} & \begin{minipage}[b]{\linewidth}\raggedright
3
\end{minipage} & \begin{minipage}[b]{\linewidth}\raggedright
4
\end{minipage} & \begin{minipage}[b]{\linewidth}\raggedright
5
\end{minipage} & \begin{minipage}[b]{\linewidth}\raggedright
6
\end{minipage} & \begin{minipage}[b]{\linewidth}\raggedright
7
\end{minipage} & \begin{minipage}[b]{\linewidth}\raggedright
8
\end{minipage} & \begin{minipage}[b]{\linewidth}\raggedright
9
\end{minipage} \\
\midrule()
\endfirsthead
\toprule()
\begin{minipage}[b]{\linewidth}\raggedright
n
\end{minipage} & \begin{minipage}[b]{\linewidth}\raggedright
2
\end{minipage} & \begin{minipage}[b]{\linewidth}\raggedright
3
\end{minipage} & \begin{minipage}[b]{\linewidth}\raggedright
4
\end{minipage} & \begin{minipage}[b]{\linewidth}\raggedright
5
\end{minipage} & \begin{minipage}[b]{\linewidth}\raggedright
6
\end{minipage} & \begin{minipage}[b]{\linewidth}\raggedright
7
\end{minipage} & \begin{minipage}[b]{\linewidth}\raggedright
8
\end{minipage} & \begin{minipage}[b]{\linewidth}\raggedright
9
\end{minipage} \\
\midrule()
\endhead
\(c_n\) & 0.743 & 1.851 & 0.954 & 1.351 & 0.993 & 1.198 & 1.005 & 1.131 \\
\(d_n\) & 0.399 & 0.994 & 0.512 & 0.844 & 0.611 & 0.857 & 0.669 & 0.872 \\
\bottomrule()
\end{longtable}

For \(n \geq 10\), \autocite{croux1992} suggests using the following prediction equations:

\begin{equation}
c_n = \begin{cases}
\frac{n}{n - 0.9}, & \quad \textrm{for odd}\; n,\\
1, & \quad \textrm{for even}\; n,
\end{cases}
\label{eq:croux-cn}
\end{equation}

\begin{equation}
d_n = \begin{cases}
\frac{n}{n + 1.4},& \quad \textrm{for odd}\; n,\\
\frac{n}{n + 3.8},& \quad \textrm{for even}\; n.
\end{cases}
\label{eq:croux-qn}
\end{equation}

\bigskip

A fast implementation of \(S_n\) and \(Q_n\) (based on \autocite{croux1992})
is available in the R package \texttt{robustbase} (see \autocite{robustbase}).
At the present moment,
the latest version of this package (0.95-0) uses\footnote{\url{https://github.com/cran/robustbase/blob/0.95-0/R/qnsn.R\#L56}}
adjusted values of \(d_n\) for \(n \leq 12\) listed in Table~\ref{tab:factors-rb}.

\begin{longtable}[]{@{}
  >{\raggedright\arraybackslash}p{(\columnwidth - 22\tabcolsep) * \real{0.0654}}
  >{\raggedright\arraybackslash}p{(\columnwidth - 22\tabcolsep) * \real{0.0935}}
  >{\raggedright\arraybackslash}p{(\columnwidth - 22\tabcolsep) * \real{0.0841}}
  >{\raggedright\arraybackslash}p{(\columnwidth - 22\tabcolsep) * \real{0.0841}}
  >{\raggedright\arraybackslash}p{(\columnwidth - 22\tabcolsep) * \real{0.0841}}
  >{\raggedright\arraybackslash}p{(\columnwidth - 22\tabcolsep) * \real{0.0841}}
  >{\raggedright\arraybackslash}p{(\columnwidth - 22\tabcolsep) * \real{0.0841}}
  >{\raggedright\arraybackslash}p{(\columnwidth - 22\tabcolsep) * \real{0.0841}}
  >{\raggedright\arraybackslash}p{(\columnwidth - 22\tabcolsep) * \real{0.0841}}
  >{\raggedright\arraybackslash}p{(\columnwidth - 22\tabcolsep) * \real{0.0841}}
  >{\raggedright\arraybackslash}p{(\columnwidth - 22\tabcolsep) * \real{0.0841}}
  >{\raggedright\arraybackslash}p{(\columnwidth - 22\tabcolsep) * \real{0.0841}}@{}}
\caption{\label{tab:factors-rb} \(Q_n\) finite-sample bias-correction factors from \texttt{robustbase\ 0.95-0}.}\tabularnewline
\toprule()
\begin{minipage}[b]{\linewidth}\raggedright
n
\end{minipage} & \begin{minipage}[b]{\linewidth}\raggedright
2
\end{minipage} & \begin{minipage}[b]{\linewidth}\raggedright
3
\end{minipage} & \begin{minipage}[b]{\linewidth}\raggedright
4
\end{minipage} & \begin{minipage}[b]{\linewidth}\raggedright
5
\end{minipage} & \begin{minipage}[b]{\linewidth}\raggedright
6
\end{minipage} & \begin{minipage}[b]{\linewidth}\raggedright
7
\end{minipage} & \begin{minipage}[b]{\linewidth}\raggedright
8
\end{minipage} & \begin{minipage}[b]{\linewidth}\raggedright
9
\end{minipage} & \begin{minipage}[b]{\linewidth}\raggedright
10
\end{minipage} & \begin{minipage}[b]{\linewidth}\raggedright
11
\end{minipage} & \begin{minipage}[b]{\linewidth}\raggedright
12
\end{minipage} \\
\midrule()
\endfirsthead
\toprule()
\begin{minipage}[b]{\linewidth}\raggedright
n
\end{minipage} & \begin{minipage}[b]{\linewidth}\raggedright
2
\end{minipage} & \begin{minipage}[b]{\linewidth}\raggedright
3
\end{minipage} & \begin{minipage}[b]{\linewidth}\raggedright
4
\end{minipage} & \begin{minipage}[b]{\linewidth}\raggedright
5
\end{minipage} & \begin{minipage}[b]{\linewidth}\raggedright
6
\end{minipage} & \begin{minipage}[b]{\linewidth}\raggedright
7
\end{minipage} & \begin{minipage}[b]{\linewidth}\raggedright
8
\end{minipage} & \begin{minipage}[b]{\linewidth}\raggedright
9
\end{minipage} & \begin{minipage}[b]{\linewidth}\raggedright
10
\end{minipage} & \begin{minipage}[b]{\linewidth}\raggedright
11
\end{minipage} & \begin{minipage}[b]{\linewidth}\raggedright
12
\end{minipage} \\
\midrule()
\endhead
\(d_n\) & 0.399356 & 0.99365 & 0.51321 & 0.84401 & 0.61220 & 0.85877 & 0.66993 & 0.87344 & 0.72014 & 0.88906 & 0.75743 \\
\bottomrule()
\end{longtable}

For \(n > 12\), the following prediction equations are used\footnote{\url{https://github.com/cran/robustbase/blob/0.95-0/R/qnsn.R\#L13}}:

\begin{equation}
d_n = \begin{cases}
\big( 1 + 1.60188 n^{-1} - 2.1284 n^{-2} - 5.172 n^{-3} \big)^{-1},& \quad \textrm{for odd}\; n,\\
\big( 1 + 3.67561 n^{-1} + 1.9654 n^{-2} + 6.987 n^{-3} - 77 n^{-4} \big)^{-1},& \quad \textrm{for even}\; n.
\end{cases}
\label{eq:robustbase-qn}
\end{equation}

\bigskip

Speaking of the finite-sample Gaussian efficiency,
\autocite{rousseeuw1993} contains only a short note about the loss of \(Q_n\) efficiency at small sample sizes
(see \autocite[p1278, before Theorem 7]{rousseeuw1993}),
and a short list of roughly approximated standardized variances of \(\MAD_n\), \(\Sn\), \(\Qn\), and \(\SD_n\)
for \(n \in \{ 10, 20, 40, 60, 80, 100, 200 \}\) under normality (see \autocite[Table 2]{rousseeuw1993}).
These data are not enough
to make a reasonable decision on a proper dispersion estimator for a small sample of the given size.

In this paper, we obtain refined values of the finite-sample bias-correction factors of \(\Sn\) and \(\Qn\)
(which are more accurate than values from \autocite{croux1992} and \autocite{robustbase})
and the corresponding finite-sample Gaussian efficiency values
using extensive Monte-Carlo simulations.

\clearpage

\hypertarget{sec:sim}{%
\section{Simulation study}\label{sec:sim}}

In this section, we perform two simulation studies in order to obtain finite-sample properties of \(\Sn\) and \(\Qn\):
the bias-correction factors for \(\Sn\) and \(\Qn\) (Section~\ref{sec:sim-factors}),
the Gaussian efficiency values (Section~\ref{sec:sim-efficiency}).

\hypertarget{sec:sim-factors}{%
\subsection{Simulation 1: Refined finite-sample bias-correction factors}\label{sec:sim-factors}}

Since \(C_n = 1/\E[\Sn(X)]\), \(D_n = 1/\E[\Qn(X)]\) and \(c_n = C_n/C_{\infty}\), \(d_n = D_n/D_{\infty}\),
the values of the finite-sample consistency constants and the corresponding bias-correction factors
for \(\Sn\) and \(\Qn\) can be obtained
by estimating the expected value of \(\Sn(x)\) and \(\Qn(x)\) using the Monte-Carlo method.
We perform the simulation according to the following scheme:

\begin{algorithm}[H]
\ForEach{$n \in \{ 2..100, \ldots, 10\,000 \}$}{
  $\textit{repetitions} \gets \textbf{when} \,\{ n \leq 100 \to 25\,000\,000;\; n > 100 \to 2\,000\,000 \}$\\
  \For{$i \gets 1..\textit{repetitions}$}{
        $x \gets \textrm{GenerateRandomSample}(\textrm{Distribution} = \mathcal{N}(0, 1),\, \textrm{SampleSize} = n)$\\
        $y_{\operatorname{S},i} \gets \Sn(x)$\\
        $y_{\operatorname{Q},i} \gets \Qn(x)$
  }
  $c_n \gets 1 \;/\; (\sum y_{\operatorname{S},i} / \textit{repetitions}) \;/\; 1.19259855312321$\\
  $d_n \gets 1 \;/\; (\sum y_{\operatorname{Q},i} / \textit{repetitions}) \;/\; 2.21914446598508$
}
\end{algorithm}

\begin{figure}[ht!]

{\centering \includegraphics{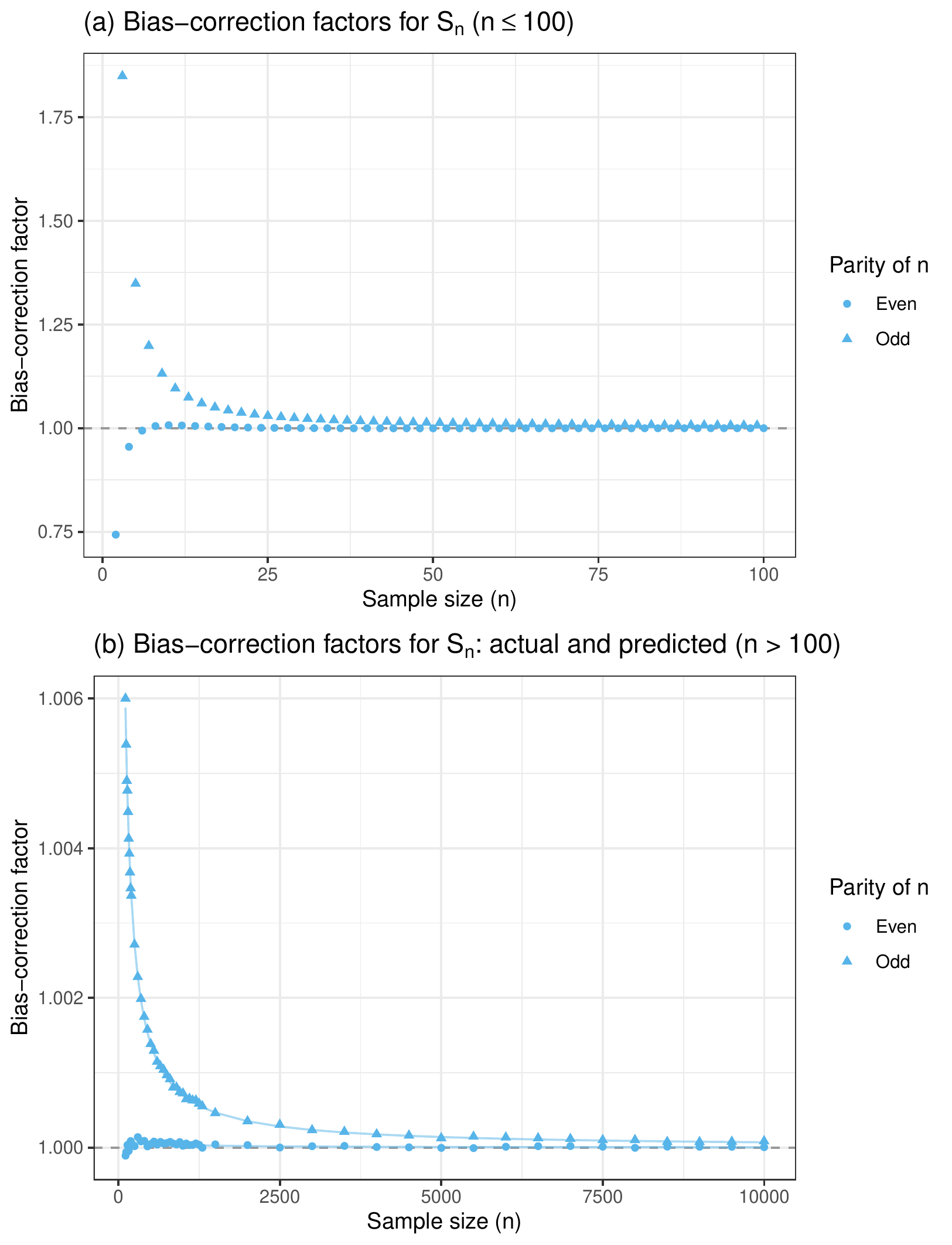} 

}

\caption{Finite-sample bias-correction factors for $\Sn$.}\label{fig:factors-sn}
\end{figure}

\begin{figure}[ht!]

{\centering \includegraphics{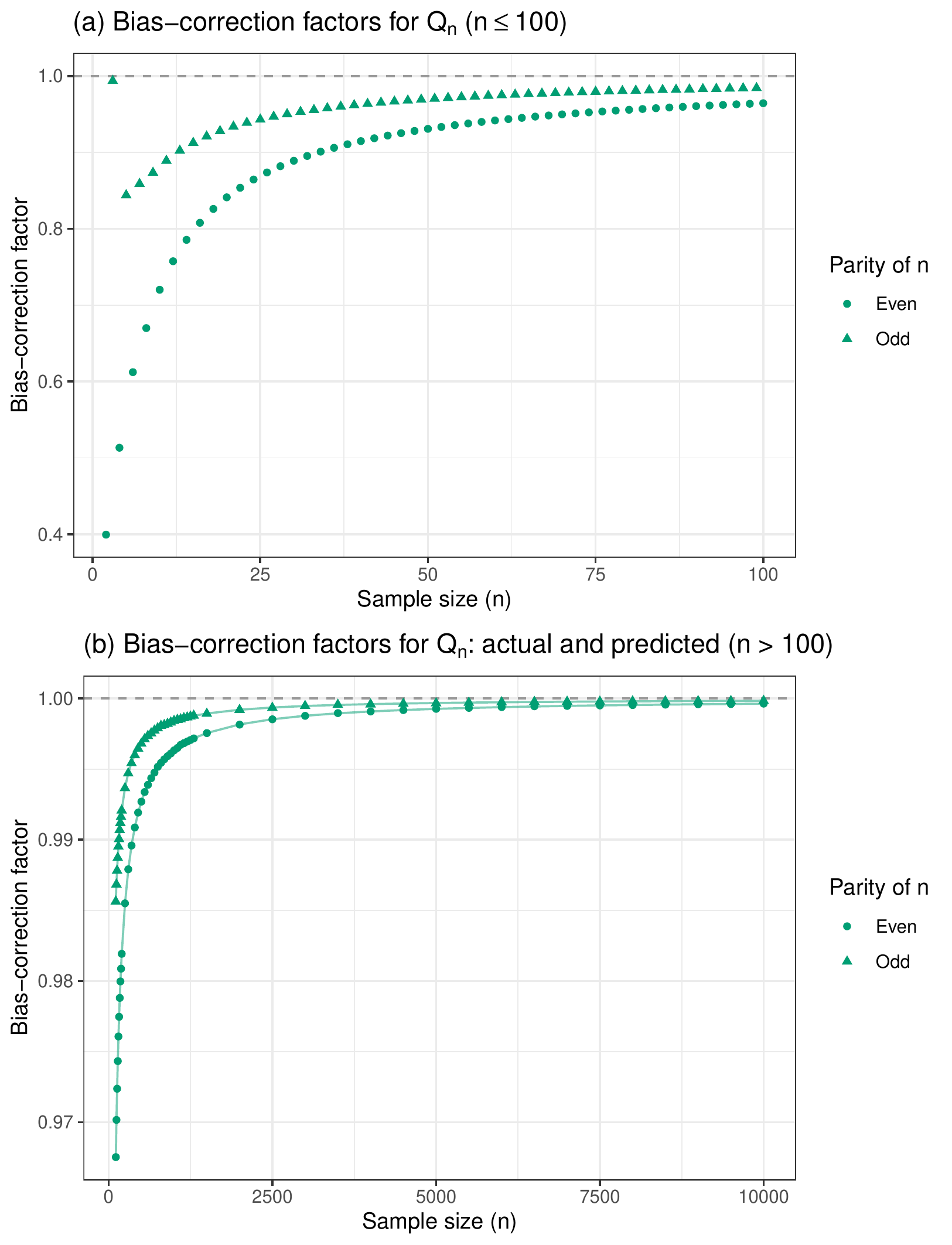} 

}

\caption{Finite-sample bias-correction factors for $\Qn$.}\label{fig:factors-qn}
\end{figure}

The estimated \(c_n,\, d_n\) values are presented in Table~\ref{tab:factors}.
The corresponding plots for \(2 \leq n \leq 100\)
are shown in Figure~\ref{fig:factors-sn}a and Figure~\ref{fig:factors-qn}a.

Following the approach from \autocite{hayes2014},
we are going to draw prediction \(c_n\) and \(d_n\) equations for \(n > 100\) as \(1 + \alpha n^{-1} + \beta n^{-2}\).
Using least squares on the values from Table~\ref{tab:factors} for \(100 < n \leq 1000\) (separately for odd and even values),
we can obtain approximated values of \(\alpha\) and \(\beta\),
which give us the following equations:

\begin{equation}
c_n = \begin{cases}
1
  +0.707 \cdot n^{-1}
  -7.181 \cdot n^{-2},& \quad \textrm{for odd}\; n,\\
1
  +0.043 \cdot n^{-1}
  -6.288 \cdot n^{-2},& \quad \textrm{for even}\; n,\\
\end{cases}
\label{eq:new-sn}
\end{equation}

\begin{equation}
d_n = \begin{cases}
1
  -1.594 \cdot n^{-1}
  +3.22 \cdot n^{-2},& \quad \textrm{for odd}\; n,\\
1
  -3.672 \cdot n^{-1}
  +11.087 \cdot n^{-2},& \quad \textrm{for even}\; n.\\
\end{cases}
\label{eq:new-qn}
\end{equation}

The actual and predicted values of \(c_n,\, d_n\)
for \(100 < n \leq 10\,000\)
are shown in Figure~\ref{fig:factors-sn}b and Figure~\ref{fig:factors-qn}b respectively.
The obtained prediction Equations~\eqref{eq:new-sn} and \eqref{eq:new-qn} look quite accurate:
the maximum observed absolute difference between the actual and predicted values
is \(\approx 0.000145\) for \(c_n\)
and \(\approx 0.000106\) for \(d_n\).

Now let us compare the obtained values from Table~\ref{tab:factors} and predicted values based on
Equation~\eqref{eq:croux-cn} and Equation~\eqref{eq:croux-qn} (the \autocite{croux1992} approach).
The maximum observed absolute difference
is \(\approx 0.007697\) for \(c_n\)
and \(\approx 0.004513\) for \(d_n\)
(observed for \(n = 25\) and \(n = 10\) respectively),
which can introduce a noticeable systematic error in the long run.

The maximum observed absolute difference between values from Table~\ref{tab:factors}
and predicted values based on Equation~\eqref{eq:robustbase-qn} (the \autocite{robustbase} approach)
is \(\approx 0.00011\) for \(d_n\)
(observed for \(n = 20\)),
which is insignificant from a practical point of view.
However, Equation~\eqref{eq:new-qn} looks more simple than Equation~\eqref{eq:robustbase-qn}.

\clearpage

\clearpage

\clearpage

\begin{table}[!h]

\caption{\label{tab:factors}Finite-sample bias-correction factors for $\Sn$ and $\Qn$.}
\centering
\begin{tabular}[t]{r|r|r|r|r|r|r|r|r|r|r|r}
\hline
n & $c_n$ & $d_n$ & n & $c_n$ & $d_n$ & n & $c_n$ & $d_n$ & n & $c_n$ & $d_n$\\
\hline
1 & - & - & 51 & 1.0127 & 0.9704 & 109 & 1.0060 & 0.9856 & 999 & 1.0007 & 0.9984\\
\hline
2 & 0.7431 & 0.3995 & 52 & 0.9996 & 0.9333 & 110 & 0.9999 & 0.9675 & 1000 & 1.0000 & 0.9963\\
\hline
3 & 1.8493 & 0.9937 & 53 & 1.0122 & 0.9714 & 119 & 1.0054 & 0.9868 & 1049 & 1.0006 & 0.9985\\
\hline
4 & 0.9550 & 0.5132 & 54 & 0.9995 & 0.9356 & 120 & 0.9999 & 0.9702 & 1050 & 1.0001 & 0.9965\\
\hline
5 & 1.3487 & 0.8440 & 55 & 1.0117 & 0.9724 & 129 & 1.0049 & 0.9878 & 1099 & 1.0007 & 0.9985\\
\hline
6 & 0.9940 & 0.6122 & 56 & 0.9995 & 0.9378 & 130 & 0.9999 & 0.9724 & 1100 & 1.0000 & 0.9967\\
\hline
7 & 1.1985 & 0.8588 & 57 & 1.0112 & 0.9733 & 139 & 1.0048 & 0.9887 & 1149 & 1.0006 & 0.9986\\
\hline
8 & 1.0050 & 0.6699 & 58 & 0.9996 & 0.9399 & 140 & 1.0000 & 0.9743 & 1150 & 1.0000 & 0.9968\\
\hline
9 & 1.1317 & 0.8734 & 59 & 1.0109 & 0.9742 & 149 & 1.0045 & 0.9895 & 1199 & 1.0006 & 0.9987\\
\hline
10 & 1.0070 & 0.7201 & 60 & 0.9996 & 0.9418 & 150 & 1.0000 & 0.9761 & 1200 & 1.0001 & 0.9970\\
\hline
11 & 1.0960 & 0.8891 & 61 & 1.0105 & 0.9750 & 159 & 1.0041 & 0.9901 & 1249 & 1.0006 & 0.9987\\
\hline
12 & 1.0063 & 0.7575 & 62 & 0.9995 & 0.9436 & 160 & 1.0000 & 0.9775 & 1250 & 1.0000 & 0.9971\\
\hline
13 & 1.0742 & 0.9023 & 63 & 1.0102 & 0.9757 & 169 & 1.0039 & 0.9907 & 1299 & 1.0006 & 0.9988\\
\hline
14 & 1.0052 & 0.7855 & 64 & 0.9996 & 0.9452 & 170 & 1.0001 & 0.9788 & 1300 & 1.0000 & 0.9972\\
\hline
15 & 1.0600 & 0.9125 & 65 & 1.0099 & 0.9764 & 179 & 1.0037 & 0.9912 & 1499 & 1.0005 & 0.9989\\
\hline
16 & 1.0039 & 0.8078 & 66 & 0.9996 & 0.9469 & 180 & 1.0000 & 0.9800 & 1500 & 1.0000 & 0.9976\\
\hline
17 & 1.0502 & 0.9210 & 67 & 1.0095 & 0.9771 & 189 & 1.0035 & 0.9916 & 1999 & 1.0004 & 0.9992\\
\hline
18 & 1.0028 & 0.8260 & 68 & 0.9996 & 0.9483 & 190 & 1.0001 & 0.9809 & 2000 & 1.0000 & 0.9982\\
\hline
19 & 1.0429 & 0.9279 & 69 & 1.0092 & 0.9778 & 199 & 1.0034 & 0.9921 & 2499 & 1.0003 & 0.9993\\
\hline
20 & 1.0021 & 0.8411 & 70 & 0.9996 & 0.9497 & 200 & 1.0000 & 0.9819 & 2500 & 1.0000 & 0.9985\\
\hline
21 & 1.0374 & 0.9338 & 71 & 1.0090 & 0.9784 & 249 & 1.0027 & 0.9937 & 2999 & 1.0002 & 0.9995\\
\hline
22 & 1.0014 & 0.8537 & 72 & 0.9996 & 0.9511 & 250 & 1.0000 & 0.9855 & 3000 & 1.0000 & 0.9988\\
\hline
23 & 1.0331 & 0.9388 & 73 & 1.0088 & 0.9789 & 299 & 1.0023 & 0.9947 & 3499 & 1.0002 & 0.9996\\
\hline
24 & 1.0009 & 0.8644 & 74 & 0.9997 & 0.9524 & 300 & 1.0001 & 0.9879 & 3500 & 1.0000 & 0.9990\\
\hline
25 & 1.0296 & 0.9431 & 75 & 1.0085 & 0.9794 & 349 & 1.0020 & 0.9954 & 3999 & 1.0002 & 0.9996\\
\hline
26 & 1.0007 & 0.8737 & 76 & 0.9997 & 0.9536 & 350 & 1.0001 & 0.9896 & 4000 & 1.0000 & 0.9991\\
\hline
27 & 1.0269 & 0.9468 & 77 & 1.0083 & 0.9800 & 399 & 1.0017 & 0.9960 & 4499 & 1.0002 & 0.9996\\
\hline
28 & 1.0004 & 0.8819 & 78 & 0.9997 & 0.9547 & 400 & 1.0001 & 0.9909 & 4500 & 1.0000 & 0.9992\\
\hline
29 & 1.0245 & 0.9501 & 79 & 1.0081 & 0.9805 & 449 & 1.0016 & 0.9965 & 4999 & 1.0001 & 0.9997\\
\hline
30 & 1.0001 & 0.8890 & 80 & 0.9996 & 0.9558 & 450 & 1.0000 & 0.9919 & 5000 & 1.0000 & 0.9993\\
\hline
31 & 1.0226 & 0.9531 & 81 & 1.0079 & 0.9809 & 499 & 1.0014 & 0.9968 & 5499 & 1.0001 & 0.9997\\
\hline
32 & 0.9999 & 0.8953 & 82 & 0.9997 & 0.9568 & 500 & 1.0000 & 0.9927 & 5500 & 1.0000 & 0.9993\\
\hline
33 & 1.0209 & 0.9556 & 83 & 1.0077 & 0.9814 & 549 & 1.0013 & 0.9971 & 5999 & 1.0001 & 0.9997\\
\hline
34 & 0.9998 & 0.9009 & 84 & 0.9997 & 0.9578 & 550 & 1.0001 & 0.9934 & 6000 & 1.0000 & 0.9994\\
\hline
35 & 1.0195 & 0.9579 & 85 & 1.0076 & 0.9818 & 599 & 1.0011 & 0.9974 & 6499 & 1.0001 & 0.9998\\
\hline
36 & 0.9997 & 0.9060 & 86 & 0.9997 & 0.9588 & 600 & 1.0000 & 0.9939 & 6500 & 1.0000 & 0.9994\\
\hline
37 & 1.0182 & 0.9600 & 87 & 1.0074 & 0.9822 & 649 & 1.0011 & 0.9975 & 6999 & 1.0001 & 0.9998\\
\hline
38 & 0.9996 & 0.9106 & 88 & 0.9997 & 0.9597 & 650 & 1.0001 & 0.9944 & 7000 & 1.0000 & 0.9995\\
\hline
39 & 1.0171 & 0.9619 & 89 & 1.0072 & 0.9825 & 699 & 1.0010 & 0.9977 & 7499 & 1.0001 & 0.9998\\
\hline
40 & 0.9997 & 0.9147 & 90 & 0.9997 & 0.9605 & 700 & 1.0001 & 0.9948 & 7500 & 1.0000 & 0.9995\\
\hline
41 & 1.0162 & 0.9636 & 91 & 1.0071 & 0.9830 & 749 & 1.0010 & 0.9979 & 7999 & 1.0001 & 0.9998\\
\hline
42 & 0.9996 & 0.9185 & 92 & 0.9997 & 0.9614 & 750 & 1.0001 & 0.9952 & 8000 & 1.0000 & 0.9995\\
\hline
43 & 1.0154 & 0.9652 & 93 & 1.0069 & 0.9833 & 799 & 1.0009 & 0.9981 & 8499 & 1.0001 & 0.9998\\
\hline
44 & 0.9996 & 0.9220 & 94 & 0.9997 & 0.9621 & 800 & 1.0001 & 0.9954 & 8500 & 1.0000 & 0.9996\\
\hline
45 & 1.0146 & 0.9667 & 95 & 1.0068 & 0.9836 & 849 & 1.0008 & 0.9981 & 8999 & 1.0001 & 0.9998\\
\hline
46 & 0.9996 & 0.9252 & 96 & 0.9998 & 0.9629 & 850 & 1.0001 & 0.9957 & 9000 & 1.0000 & 0.9996\\
\hline
47 & 1.0139 & 0.9680 & 97 & 1.0067 & 0.9840 & 899 & 1.0008 & 0.9982 & 9499 & 1.0001 & 0.9998\\
\hline
48 & 0.9995 & 0.9281 & 98 & 0.9998 & 0.9636 & 900 & 1.0000 & 0.9959 & 9500 & 1.0000 & 0.9996\\
\hline
49 & 1.0133 & 0.9692 & 99 & 1.0065 & 0.9843 & 949 & 1.0007 & 0.9983 & 9999 & 1.0001 & 0.9998\\
\hline
50 & 0.9995 & 0.9308 & 100 & 0.9998 & 0.9644 & 950 & 1.0001 & 0.9961 & 10000 & 1.0000 & 0.9996\\
\hline
\end{tabular}
\end{table}

\clearpage

\hypertarget{sec:sim-efficiency}{%
\subsection{Simulation 2: Finite-sample Gaussian efficiency}\label{sec:sim-efficiency}}

In this simulation study, we evaluate the finite-sample Gaussian efficiency of \(\MAD_n\), \(\Sn\), and \(\Qn\).
As for the baseline, we consider the unbiased standard deviation \(\SD_n\) of the normal distribution:

\[
\SD_n(X) = \sqrt{\frac{1}{n} \sum_{i=1}^n (X_i - \bar{X})^2} \bigg/ c_4(n), \quad
c_4(n) = \sqrt{\frac{2}{n-1}}\frac{\Gamma(\frac{n}{2})}{\Gamma(\frac{n-1}{2})}.
\]

For the \(\MAD_n\), we use bias-correction factors from \autocite{akinshin2022madfactors}.
For \(\Sn\) and \(\Qn\), we use freshly obtained bias-correction factors from Table~\ref{tab:factors} and
Equations~\eqref{eq:new-sn} and \eqref{eq:new-qn}.

For an unbiased scale estimator \(T_n\), the Gaussian efficiency is defined as follows:

\begin{equation}
e(T_n) = \frac{\V[\SD_n]}{\V[T_n]}.
\label{eq:dispersion-efficiency}
\end{equation}

We also consider the concept of the \emph{standardized asymptotic variance} of a scale estimator which was
proposed in \autocite{daniell1920} and advocated in \autocite[p1276]{rousseeuw1993}, \autocite[p502]{bickel1976}, and \autocite[p3]{huber2009}:

\begin{equation}
\V_s[T_n] = \frac{n \cdot \V[T_n]}{\E[T_n]^2}
\label{eq:svar}
\end{equation}

We suggest overriding Equation~\eqref{eq:dispersion-efficiency} using \(\V_s\):

\begin{equation}
e(T_n) = \frac{\V_s[\SD_n]}{\V_s[T_n]}.
\label{eq:dispersion-efficiency2}
\end{equation}

Equations~\eqref{eq:dispersion-efficiency} and \eqref{eq:dispersion-efficiency2} are equivalent for unbiased estimators
since \(\E[T_n] = 1\).
However, we operate only with approximations of the consistency constants
for the \(\MAD_n\), \(\Sn\) and \(\Qn\) (see Table~\ref{tab:factors}).
The difference between the actual and approximated values
of the consistency constants is almost negligible in practice,
but it still introduces minor errors.
In order to slightly improve the accuracy of the estimated Gaussian efficiency values, we prefer using
Equation~\eqref{eq:dispersion-efficiency2} in our calculations.

We perform the simulation using the Monte-Carlo method according to the following scheme:

\begin{algorithm}[H]
\ForEach{$n \in \{  2..100, \ldots, 100\,000 \}$}{
  $\textit{repetitions} \gets \textbf{when} \,\{ n \leq 100 \to 10\,000\,000;\; n > 100 \to 2\,000\,000 \}$\\
  \For{$i \gets 1..\textit{repetitions}$}{
    $x \gets \textrm{GenerateRandomSample}(\textrm{Distribution} = \mathcal{N}(0, 1),\, \textrm{SampleSize} = n)$\\
    $y_{\SD,i} \gets \SD_n(x)$\\
    $y_{\MAD,i} \gets \MAD_n(x)$\\
    $y_{\operatorname{S},i} \gets \Sn(x)$\\
    $y_{\operatorname{Q},i} \gets \Qn(x)$\\
  }
  $e(\MAD_n) \gets \V_s(y_{\SD,\{i\}}) \;/\; \V_s(y_{\MAD,\{i\}})$\\
  $e(\Sn) \gets \V_s(y_{\SD,\{i\}}) \;/\; \V_s(y_{\operatorname{S},\{i\}})$\\
  $e(\Qn) \gets \V_s(y_{\SD,\{i\}}) \;/\; \V_s(y_{\operatorname{Q},\{i\}})$\\
}
\end{algorithm}

\begin{figure}[ht!]

{\centering \includegraphics{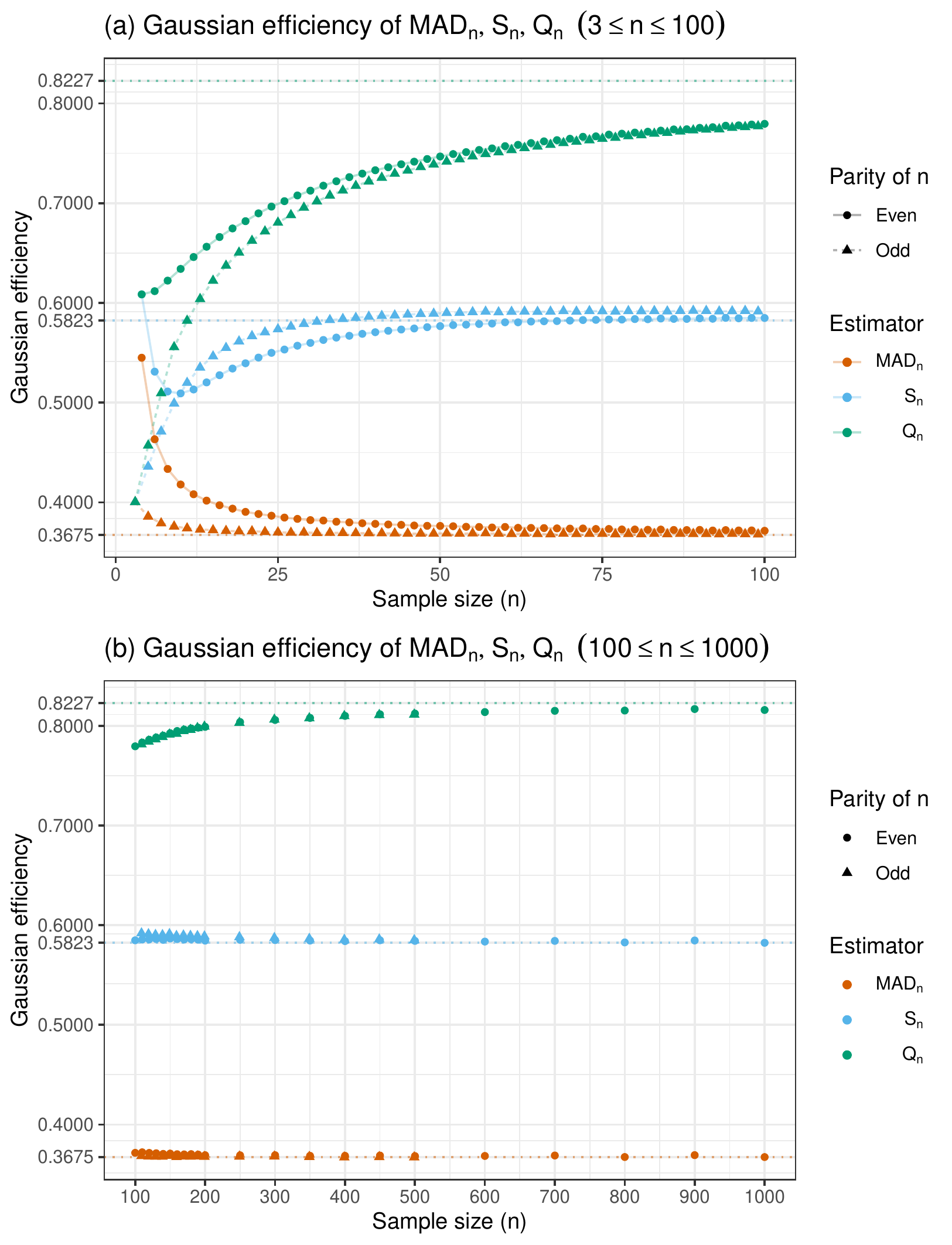} 

}

\caption{Finite-sample Gaussian efficiency of $\MAD_n$, $\Sn$, $\Qn$.}\label{fig:fig-efficiency}
\end{figure}

The estimated Gaussian efficiency values are presented in Table~\ref{tab:tab-efficiency}.
The corresponding plots for \(3 \leq n \leq 100\) and \(100 \leq n \leq 1\;000\)
are shown in Figure~\ref{fig:fig-efficiency}.

As we can see, the actual finite-sample Gaussian efficiency of \(\Sn\) and \(\Qn\) on small samples are noticeable smaller than
the corresponding asymptotic values.
For example, \(e(\operatorname{S}_7) \approx 47\%\) and \(e(\operatorname{Q}_7) \approx 51\%\)
compared to \(\approx 58\%\) and \(\approx 82\%\) in the asymptotic case.
However, the finite-sample \(\Sn\) and \(\Qn\) are still more efficient than \(\MAD_n\).

\clearpage

\clearpage

\begin{table}[!h]

\caption{\label{tab:tab-efficiency}Finite-sample Gaussian efficiency of $\MAD_n$, $\Sn$, $\Qn$.}
\centering
\begin{tabular}[t]{r|r|r|r|r|r|r|r|r|r|r|r}
\hline
n & $\MAD_n$ & $\Sn$ & $\Qn$ & n & $\MAD_n$ & $\Sn$ & $\Qn$ & n & $\MAD_n$ & $\Sn$ & $\Qn$\\
\hline
1 & - & - & - & 51 & 0.3690 & 0.5900 & 0.7417 & 109 & 0.3685 & 0.5919 & 0.7815\\
\hline
2 & 1.0000 & 1.0000 & 1.0000 & 52 & 0.3761 & 0.5777 & 0.7492 & 110 & 0.3722 & 0.5856 & 0.7833\\
\hline
3 & 0.4005 & 0.4005 & 0.4005 & 53 & 0.3687 & 0.5901 & 0.7440 & 119 & 0.3680 & 0.5910 & 0.7844\\
\hline
4 & 0.5451 & 0.6085 & 0.6085 & 54 & 0.3756 & 0.5782 & 0.7511 & 120 & 0.3716 & 0.5857 & 0.7861\\
\hline
5 & 0.3859 & 0.4360 & 0.4571 & 55 & 0.3690 & 0.5908 & 0.7469 & 129 & 0.3679 & 0.5903 & 0.7865\\
\hline
6 & 0.4633 & 0.5311 & 0.6118 & 56 & 0.3754 & 0.5788 & 0.7533 & 130 & 0.3712 & 0.5861 & 0.7884\\
\hline
7 & 0.3791 & 0.4713 & 0.5096 & 57 & 0.3691 & 0.5913 & 0.7493 & 139 & 0.3678 & 0.5901 & 0.7889\\
\hline
8 & 0.4336 & 0.5112 & 0.6223 & 58 & 0.3749 & 0.5793 & 0.7549 & 140 & 0.3706 & 0.5853 & 0.7902\\
\hline
9 & 0.3760 & 0.4992 & 0.5557 & 59 & 0.3688 & 0.5907 & 0.7510 & 149 & 0.3688 & 0.5911 & 0.7913\\
\hline
10 & 0.4180 & 0.5093 & 0.6341 & 60 & 0.3754 & 0.5806 & 0.7571 & 150 & 0.3709 & 0.5866 & 0.7926\\
\hline
11 & 0.3741 & 0.5199 & 0.5822 & 61 & 0.3685 & 0.5911 & 0.7533 & 159 & 0.3675 & 0.5892 & 0.7920\\
\hline
12 & 0.4082 & 0.5132 & 0.6460 & 62 & 0.3746 & 0.5805 & 0.7584 & 160 & 0.3703 & 0.5861 & 0.7948\\
\hline
13 & 0.3728 & 0.5352 & 0.6039 & 63 & 0.3689 & 0.5914 & 0.7553 & 169 & 0.3680 & 0.5898 & 0.7948\\
\hline
14 & 0.4018 & 0.5203 & 0.6563 & 64 & 0.3745 & 0.5813 & 0.7602 & 170 & 0.3701 & 0.5857 & 0.7963\\
\hline
15 & 0.3723 & 0.5464 & 0.6223 & 65 & 0.3685 & 0.5911 & 0.7566 & 179 & 0.3679 & 0.5894 & 0.7960\\
\hline
16 & 0.3972 & 0.5275 & 0.6660 & 66 & 0.3743 & 0.5816 & 0.7619 & 180 & 0.3703 & 0.5859 & 0.7974\\
\hline
17 & 0.3716 & 0.5549 & 0.6374 & 67 & 0.3681 & 0.5913 & 0.7584 & 189 & 0.3682 & 0.5891 & 0.7978\\
\hline
18 & 0.3938 & 0.5342 & 0.6746 & 68 & 0.3740 & 0.5818 & 0.7631 & 190 & 0.3699 & 0.5855 & 0.7981\\
\hline
19 & 0.3710 & 0.5615 & 0.6505 & 69 & 0.3687 & 0.5918 & 0.7602 & 199 & 0.3677 & 0.5888 & 0.7992\\
\hline
20 & 0.3905 & 0.5395 & 0.6819 & 70 & 0.3738 & 0.5823 & 0.7645 & 200 & 0.3693 & 0.5845 & 0.7990\\
\hline
21 & 0.3712 & 0.5668 & 0.6623 & 71 & 0.3684 & 0.5914 & 0.7617 & 249 & 0.3679 & 0.5880 & 0.8030\\
\hline
22 & 0.3884 & 0.5452 & 0.6898 & 72 & 0.3739 & 0.5828 & 0.7665 & 250 & 0.3692 & 0.5853 & 0.8038\\
\hline
23 & 0.3705 & 0.5706 & 0.6717 & 73 & 0.3685 & 0.5919 & 0.7631 & 299 & 0.3680 & 0.5872 & 0.8063\\
\hline
24 & 0.3867 & 0.5500 & 0.6965 & 74 & 0.3734 & 0.5830 & 0.7669 & 300 & 0.3693 & 0.5851 & 0.8059\\
\hline
25 & 0.3701 & 0.5737 & 0.6805 & 75 & 0.3686 & 0.5918 & 0.7645 & 349 & 0.3676 & 0.5864 & 0.8076\\
\hline
26 & 0.3848 & 0.5533 & 0.7020 & 76 & 0.3734 & 0.5834 & 0.7687 & 350 & 0.3686 & 0.5846 & 0.8079\\
\hline
27 & 0.3697 & 0.5762 & 0.6880 & 77 & 0.3682 & 0.5914 & 0.7655 & 399 & 0.3671 & 0.5854 & 0.8098\\
\hline
28 & 0.3835 & 0.5571 & 0.7077 & 78 & 0.3729 & 0.5833 & 0.7696 & 400 & 0.3686 & 0.5840 & 0.8100\\
\hline
29 & 0.3700 & 0.5789 & 0.6955 & 79 & 0.3684 & 0.5919 & 0.7672 & 449 & 0.3674 & 0.5858 & 0.8109\\
\hline
30 & 0.3823 & 0.5599 & 0.7125 & 80 & 0.3729 & 0.5838 & 0.7707 & 450 & 0.3689 & 0.5849 & 0.8116\\
\hline
31 & 0.3695 & 0.5811 & 0.7019 & 81 & 0.3683 & 0.5917 & 0.7682 & 499 & 0.3672 & 0.5851 & 0.8114\\
\hline
32 & 0.3819 & 0.5629 & 0.7175 & 82 & 0.3726 & 0.5836 & 0.7716 & 500 & 0.3684 & 0.5842 & 0.8123\\
\hline
33 & 0.3696 & 0.5829 & 0.7076 & 83 & 0.3685 & 0.5919 & 0.7696 & 600 & 0.3685 & 0.5836 & 0.8138\\
\hline
34 & 0.3807 & 0.5652 & 0.7219 & 84 & 0.3728 & 0.5843 & 0.7727 & 700 & 0.3690 & 0.5842 & 0.8151\\
\hline
35 & 0.3695 & 0.5840 & 0.7126 & 85 & 0.3682 & 0.5915 & 0.7705 & 800 & 0.3674 & 0.5827 & 0.8153\\
\hline
36 & 0.3800 & 0.5671 & 0.7260 & 86 & 0.3724 & 0.5840 & 0.7739 & 900 & 0.3693 & 0.5847 & 0.8168\\
\hline
37 & 0.3693 & 0.5851 & 0.7173 & 87 & 0.3685 & 0.5920 & 0.7719 & 1000 & 0.3674 & 0.5823 & 0.8159\\
\hline
38 & 0.3792 & 0.5690 & 0.7296 & 88 & 0.3723 & 0.5840 & 0.7740 & 1500 & 0.3682 & 0.5832 & 0.8182\\
\hline
39 & 0.3692 & 0.5862 & 0.7218 & 89 & 0.3685 & 0.5922 & 0.7730 & 2000 & 0.3684 & 0.5824 & 0.8200\\
\hline
40 & 0.3786 & 0.5706 & 0.7330 & 90 & 0.3725 & 0.5845 & 0.7755 & 3000 & 0.3681 & 0.5818 & 0.8192\\
\hline
41 & 0.3691 & 0.5870 & 0.7255 & 91 & 0.3684 & 0.5920 & 0.7737 & 4000 & 0.3675 & 0.5817 & 0.8193\\
\hline
42 & 0.3779 & 0.5720 & 0.7360 & 92 & 0.3718 & 0.5841 & 0.7761 & 5000 & 0.3673 & 0.5809 & 0.8194\\
\hline
43 & 0.3690 & 0.5875 & 0.7295 & 93 & 0.3681 & 0.5915 & 0.7741 & 6000 & 0.3683 & 0.5821 & 0.8207\\
\hline
44 & 0.3775 & 0.5733 & 0.7389 & 94 & 0.3723 & 0.5850 & 0.7776 & 7000 & 0.3674 & 0.5820 & 0.8205\\
\hline
45 & 0.3688 & 0.5883 & 0.7326 & 95 & 0.3686 & 0.5917 & 0.7755 & 8000 & 0.3677 & 0.5820 & 0.8210\\
\hline
46 & 0.3769 & 0.5743 & 0.7415 & 96 & 0.3719 & 0.5846 & 0.7779 & 9000 & 0.3666 & 0.5813 & 0.8196\\
\hline
47 & 0.3689 & 0.5889 & 0.7362 & 97 & 0.3683 & 0.5921 & 0.7763 & 10000 & 0.3678 & 0.5821 & 0.8207\\
\hline
48 & 0.3766 & 0.5757 & 0.7442 & 98 & 0.3720 & 0.5849 & 0.7788 & 25000 & 0.3679 & 0.5819 & 0.8204\\
\hline
49 & 0.3688 & 0.5893 & 0.7388 & 99 & 0.3681 & 0.5915 & 0.7771 & 50000 & 0.3681 & 0.5830 & 0.8214\\
\hline
50 & 0.3765 & 0.5767 & 0.7467 & 100 & 0.3716 & 0.5848 & 0.7795 & 100000 & 0.3680 & 0.5821 & 0.8211\\
\hline
\end{tabular}
\end{table}

\clearpage

\hypertarget{sec:summary}{%
\section{Summary}\label{sec:summary}}

In this paper, we revised the finite-sample properties of the Rousseeuw-Croux \(S_n\) and \(Q_n\) scale estimators.

Firstly, we obtained refined finite-samples bias-correction factors for \(S_n\) and \(Q_n\),
which allows using them as robust consistent estimators for the standard deviation under normality.
The new factor values are described by Table~\ref{tab:factors} and
prediction Equations~\eqref{eq:new-sn} and \eqref{eq:new-qn}.
These values are more accurate than the original values from \autocite{croux1992}.
We recommend using the presented values in order to reduce the systematic bias of obtained estimations.

Secondly, we evaluated the finite-sample Gaussian efficiency values for \(S_n\) and \(Q_n\)
(Table~\ref{tab:tab-efficiency}),
which are noticeably smaller than the corresponding asymptotic values.
The knowledge of the actual finite-sample efficiency
is essential for the reasonable decision of a proper scale estimator.
The usage of asymptotic values in finite cases may also lead to poor choice of the target sample size,
which would give insufficient statistical efficiency.

\hypertarget{disclosure-statement}{%
\section*{Disclosure statement}\label{disclosure-statement}}
\addcontentsline{toc}{section}{Disclosure statement}

The author reports there are no competing interests to declare.

\hypertarget{data-and-source-code-availability}{%
\section*{Data and source code availability}\label{data-and-source-code-availability}}
\addcontentsline{toc}{section}{Data and source code availability}

The source code of this paper, the source code of all simulations,
and the simulation results are available on GitHub:
\url{https://github.com/AndreyAkinshin/paper-frc}.

\hypertarget{acknowledgments}{%
\section*{Acknowledgments}\label{acknowledgments}}
\addcontentsline{toc}{section}{Acknowledgments}

The author thanks Ivan Pashchenko for valuable discussions.

\clearpage

\hypertarget{appendix-appendix}{%
\appendix}

\hypertarget{sec:refimpl}{%
\section{Reference implementation}\label{sec:refimpl}}

In our R reference implementation, we reuse the existing solution from the package \texttt{robustbase} (\autocite{robustbase})
based on the fast \(\BigO(n \log n)\) algorithm from \autocite{croux1992}
and just redefine the bias-correction factors:

\begin{Shaded}
\begin{Highlighting}[]
\FunctionTok{library}\NormalTok{(robustbase)}

\NormalTok{Sn }\OtherTok{\textless{}{-}} \ControlFlowTok{function}\NormalTok{(x) \{}
\NormalTok{  x }\OtherTok{\textless{}{-}}\NormalTok{ x[}\SpecialCharTok{!}\FunctionTok{is.na}\NormalTok{(x)]}
\NormalTok{  n }\OtherTok{\textless{}{-}} \FunctionTok{length}\NormalTok{(x)}
  \ControlFlowTok{if}\NormalTok{ (n }\SpecialCharTok{==} \DecValTok{0}\NormalTok{) }\FunctionTok{return}\NormalTok{(}\ConstantTok{NA}\NormalTok{)}
\NormalTok{  factors }\OtherTok{\textless{}{-}}
    \FunctionTok{c}\NormalTok{(    }\ConstantTok{NA}\NormalTok{, }\FloatTok{0.7430}\NormalTok{, }\FloatTok{1.8498}\NormalTok{, }\FloatTok{0.9551}\NormalTok{, }\FloatTok{1.3486}\NormalTok{, }\FloatTok{0.9941}\NormalTok{, }\FloatTok{1.1983}\NormalTok{, }\FloatTok{1.0050}\NormalTok{, }\FloatTok{1.1318}\NormalTok{, }\FloatTok{1.0069}\NormalTok{,}
      \FloatTok{1.0959}\NormalTok{, }\FloatTok{1.0063}\NormalTok{, }\FloatTok{1.0742}\NormalTok{, }\FloatTok{1.0051}\NormalTok{, }\FloatTok{1.0601}\NormalTok{, }\FloatTok{1.0038}\NormalTok{, }\FloatTok{1.0501}\NormalTok{, }\FloatTok{1.0028}\NormalTok{, }\FloatTok{1.0430}\NormalTok{, }\FloatTok{1.0022}\NormalTok{,}
      \FloatTok{1.0374}\NormalTok{, }\FloatTok{1.0014}\NormalTok{, }\FloatTok{1.0331}\NormalTok{, }\FloatTok{1.0009}\NormalTok{, }\FloatTok{1.0297}\NormalTok{, }\FloatTok{1.0007}\NormalTok{, }\FloatTok{1.0269}\NormalTok{, }\FloatTok{1.0004}\NormalTok{, }\FloatTok{1.0245}\NormalTok{, }\FloatTok{1.0001}\NormalTok{,}
      \FloatTok{1.0226}\NormalTok{, }\FloatTok{0.9999}\NormalTok{, }\FloatTok{1.0209}\NormalTok{, }\FloatTok{0.9997}\NormalTok{, }\FloatTok{1.0195}\NormalTok{, }\FloatTok{0.9998}\NormalTok{, }\FloatTok{1.0183}\NormalTok{, }\FloatTok{0.9996}\NormalTok{, }\FloatTok{1.0172}\NormalTok{, }\FloatTok{0.9997}\NormalTok{,}
      \FloatTok{1.0162}\NormalTok{, }\FloatTok{0.9996}\NormalTok{, }\FloatTok{1.0154}\NormalTok{, }\FloatTok{0.9996}\NormalTok{, }\FloatTok{1.0146}\NormalTok{, }\FloatTok{0.9996}\NormalTok{, }\FloatTok{1.0139}\NormalTok{, }\FloatTok{0.9995}\NormalTok{, }\FloatTok{1.0132}\NormalTok{, }\FloatTok{0.9995}\NormalTok{,}
      \FloatTok{1.0126}\NormalTok{, }\FloatTok{0.9995}\NormalTok{, }\FloatTok{1.0123}\NormalTok{, }\FloatTok{0.9995}\NormalTok{, }\FloatTok{1.0117}\NormalTok{, }\FloatTok{0.9995}\NormalTok{, }\FloatTok{1.0113}\NormalTok{, }\FloatTok{0.9996}\NormalTok{, }\FloatTok{1.0109}\NormalTok{, }\FloatTok{0.9996}\NormalTok{,}
      \FloatTok{1.0105}\NormalTok{, }\FloatTok{0.9995}\NormalTok{, }\FloatTok{1.0102}\NormalTok{, }\FloatTok{0.9996}\NormalTok{, }\FloatTok{1.0099}\NormalTok{, }\FloatTok{0.9997}\NormalTok{, }\FloatTok{1.0095}\NormalTok{, }\FloatTok{0.9996}\NormalTok{, }\FloatTok{1.0092}\NormalTok{, }\FloatTok{0.9997}\NormalTok{,}
      \FloatTok{1.0090}\NormalTok{, }\FloatTok{0.9997}\NormalTok{, }\FloatTok{1.0088}\NormalTok{, }\FloatTok{0.9996}\NormalTok{, }\FloatTok{1.0085}\NormalTok{, }\FloatTok{0.9997}\NormalTok{, }\FloatTok{1.0084}\NormalTok{, }\FloatTok{0.9997}\NormalTok{, }\FloatTok{1.0081}\NormalTok{, }\FloatTok{0.9997}\NormalTok{,}
      \FloatTok{1.0079}\NormalTok{, }\FloatTok{0.9997}\NormalTok{, }\FloatTok{1.0076}\NormalTok{, }\FloatTok{0.9997}\NormalTok{, }\FloatTok{1.0076}\NormalTok{, }\FloatTok{0.9997}\NormalTok{, }\FloatTok{1.0074}\NormalTok{, }\FloatTok{0.9997}\NormalTok{, }\FloatTok{1.0072}\NormalTok{, }\FloatTok{0.9997}\NormalTok{,}
      \FloatTok{1.0070}\NormalTok{, }\FloatTok{0.9997}\NormalTok{, }\FloatTok{1.0069}\NormalTok{, }\FloatTok{0.9997}\NormalTok{, }\FloatTok{1.0067}\NormalTok{, }\FloatTok{0.9998}\NormalTok{, }\FloatTok{1.0066}\NormalTok{, }\FloatTok{0.9997}\NormalTok{, }\FloatTok{1.0065}\NormalTok{, }\FloatTok{0.9998}\NormalTok{)}
\NormalTok{  predict\_factor }\OtherTok{\textless{}{-}} \ControlFlowTok{function}\NormalTok{(n) \{}
    \FunctionTok{ifelse}\NormalTok{(n }\SpecialCharTok{\%\%} \DecValTok{2} \SpecialCharTok{==} \DecValTok{1}\NormalTok{, }\DecValTok{1} \SpecialCharTok{+} \FloatTok{0.707} \SpecialCharTok{/}\NormalTok{ n }\SpecialCharTok{{-}} \FloatTok{7.181} \SpecialCharTok{/}\NormalTok{ n}\SpecialCharTok{\^{}}\DecValTok{2}\NormalTok{, }\DecValTok{1} \SpecialCharTok{+} \FloatTok{0.043} \SpecialCharTok{/}\NormalTok{ n }\SpecialCharTok{{-}} \FloatTok{6.288} \SpecialCharTok{/}\NormalTok{ n}\SpecialCharTok{\^{}}\DecValTok{2}\NormalTok{)}
\NormalTok{  \}}
\NormalTok{  factor }\OtherTok{\textless{}{-}} \ControlFlowTok{if}\NormalTok{ (n }\SpecialCharTok{\textless{}=} \DecValTok{100}\NormalTok{) factors[n] }\ControlFlowTok{else} \FunctionTok{predict\_factor}\NormalTok{(n)}
\NormalTok{  constant }\OtherTok{\textless{}{-}} \FloatTok{1.19259855312321} \SpecialCharTok{*}\NormalTok{ factor}
\NormalTok{  robustbase}\SpecialCharTok{::}\FunctionTok{Sn}\NormalTok{(x, constant)}
\NormalTok{\}}

\NormalTok{Qn }\OtherTok{\textless{}{-}} \ControlFlowTok{function}\NormalTok{(x, }\AttributeTok{constant =} \ConstantTok{NULL}\NormalTok{) \{}
\NormalTok{  x }\OtherTok{\textless{}{-}}\NormalTok{ x[}\SpecialCharTok{!}\FunctionTok{is.na}\NormalTok{(x)]}
\NormalTok{  n }\OtherTok{\textless{}{-}} \FunctionTok{length}\NormalTok{(x)}
  \ControlFlowTok{if}\NormalTok{ (n }\SpecialCharTok{==} \DecValTok{0}\NormalTok{) }\FunctionTok{return}\NormalTok{(}\ConstantTok{NA}\NormalTok{)}
\NormalTok{  factors }\OtherTok{\textless{}{-}}
    \FunctionTok{c}\NormalTok{(    }\ConstantTok{NA}\NormalTok{, }\FloatTok{0.3995}\NormalTok{, }\FloatTok{0.9939}\NormalTok{, }\FloatTok{0.5133}\NormalTok{, }\FloatTok{0.8441}\NormalTok{, }\FloatTok{0.6122}\NormalTok{, }\FloatTok{0.8589}\NormalTok{, }\FloatTok{0.6700}\NormalTok{, }\FloatTok{0.8736}\NormalTok{, }\FloatTok{0.7201}\NormalTok{,}
      \FloatTok{0.8890}\NormalTok{, }\FloatTok{0.7575}\NormalTok{, }\FloatTok{0.9023}\NormalTok{, }\FloatTok{0.7855}\NormalTok{, }\FloatTok{0.9125}\NormalTok{, }\FloatTok{0.8078}\NormalTok{, }\FloatTok{0.9211}\NormalTok{, }\FloatTok{0.8260}\NormalTok{, }\FloatTok{0.9279}\NormalTok{, }\FloatTok{0.8410}\NormalTok{,}
      \FloatTok{0.9338}\NormalTok{, }\FloatTok{0.8537}\NormalTok{, }\FloatTok{0.9389}\NormalTok{, }\FloatTok{0.8644}\NormalTok{, }\FloatTok{0.9430}\NormalTok{, }\FloatTok{0.8737}\NormalTok{, }\FloatTok{0.9468}\NormalTok{, }\FloatTok{0.8819}\NormalTok{, }\FloatTok{0.9501}\NormalTok{, }\FloatTok{0.8890}\NormalTok{,}
      \FloatTok{0.9530}\NormalTok{, }\FloatTok{0.8953}\NormalTok{, }\FloatTok{0.9557}\NormalTok{, }\FloatTok{0.9010}\NormalTok{, }\FloatTok{0.9579}\NormalTok{, }\FloatTok{0.9060}\NormalTok{, }\FloatTok{0.9600}\NormalTok{, }\FloatTok{0.9106}\NormalTok{, }\FloatTok{0.9619}\NormalTok{, }\FloatTok{0.9148}\NormalTok{,}
      \FloatTok{0.9636}\NormalTok{, }\FloatTok{0.9185}\NormalTok{, }\FloatTok{0.9652}\NormalTok{, }\FloatTok{0.9220}\NormalTok{, }\FloatTok{0.9667}\NormalTok{, }\FloatTok{0.9252}\NormalTok{, }\FloatTok{0.9680}\NormalTok{, }\FloatTok{0.9281}\NormalTok{, }\FloatTok{0.9692}\NormalTok{, }\FloatTok{0.9309}\NormalTok{,}
      \FloatTok{0.9704}\NormalTok{, }\FloatTok{0.9333}\NormalTok{, }\FloatTok{0.9715}\NormalTok{, }\FloatTok{0.9357}\NormalTok{, }\FloatTok{0.9724}\NormalTok{, }\FloatTok{0.9378}\NormalTok{, }\FloatTok{0.9733}\NormalTok{, }\FloatTok{0.9399}\NormalTok{, }\FloatTok{0.9742}\NormalTok{, }\FloatTok{0.9418}\NormalTok{,}
      \FloatTok{0.9750}\NormalTok{, }\FloatTok{0.9435}\NormalTok{, }\FloatTok{0.9757}\NormalTok{, }\FloatTok{0.9453}\NormalTok{, }\FloatTok{0.9765}\NormalTok{, }\FloatTok{0.9469}\NormalTok{, }\FloatTok{0.9771}\NormalTok{, }\FloatTok{0.9484}\NormalTok{, }\FloatTok{0.9777}\NormalTok{, }\FloatTok{0.9498}\NormalTok{,}
      \FloatTok{0.9784}\NormalTok{, }\FloatTok{0.9511}\NormalTok{, }\FloatTok{0.9789}\NormalTok{, }\FloatTok{0.9523}\NormalTok{, }\FloatTok{0.9794}\NormalTok{, }\FloatTok{0.9536}\NormalTok{, }\FloatTok{0.9800}\NormalTok{, }\FloatTok{0.9547}\NormalTok{, }\FloatTok{0.9805}\NormalTok{, }\FloatTok{0.9558}\NormalTok{,}
      \FloatTok{0.9809}\NormalTok{, }\FloatTok{0.9568}\NormalTok{, }\FloatTok{0.9814}\NormalTok{, }\FloatTok{0.9578}\NormalTok{, }\FloatTok{0.9818}\NormalTok{, }\FloatTok{0.9587}\NormalTok{, }\FloatTok{0.9822}\NormalTok{, }\FloatTok{0.9597}\NormalTok{, }\FloatTok{0.9826}\NormalTok{, }\FloatTok{0.9605}\NormalTok{,}
      \FloatTok{0.9829}\NormalTok{, }\FloatTok{0.9614}\NormalTok{, }\FloatTok{0.9833}\NormalTok{, }\FloatTok{0.9621}\NormalTok{, }\FloatTok{0.9836}\NormalTok{, }\FloatTok{0.9629}\NormalTok{, }\FloatTok{0.9840}\NormalTok{, }\FloatTok{0.9636}\NormalTok{, }\FloatTok{0.9843}\NormalTok{, }\FloatTok{0.9644}\NormalTok{)}
\NormalTok{  predict\_factor }\OtherTok{\textless{}{-}} \ControlFlowTok{function}\NormalTok{(n) \{}
    \FunctionTok{ifelse}\NormalTok{(n }\SpecialCharTok{\%\%} \DecValTok{2} \SpecialCharTok{==} \DecValTok{1}\NormalTok{, }\DecValTok{1} \SpecialCharTok{{-}} \FloatTok{1.594} \SpecialCharTok{/}\NormalTok{ n }\SpecialCharTok{+} \FloatTok{3.22} \SpecialCharTok{/}\NormalTok{ n}\SpecialCharTok{\^{}}\DecValTok{2}\NormalTok{, }\DecValTok{1} \SpecialCharTok{{-}} \FloatTok{3.672} \SpecialCharTok{/}\NormalTok{ n }\SpecialCharTok{+} \FloatTok{11.087} \SpecialCharTok{/}\NormalTok{ n}\SpecialCharTok{\^{}}\DecValTok{2}\NormalTok{)}
\NormalTok{  \}}
\NormalTok{  factor }\OtherTok{\textless{}{-}} \ControlFlowTok{if}\NormalTok{ (n }\SpecialCharTok{\textless{}=} \DecValTok{100}\NormalTok{) factors[n] }\ControlFlowTok{else} \FunctionTok{predict\_factor}\NormalTok{(n)}
\NormalTok{  constant }\OtherTok{\textless{}{-}}\NormalTok{ (}\DecValTok{1} \SpecialCharTok{/}\NormalTok{ (}\FunctionTok{sqrt}\NormalTok{(}\DecValTok{2}\NormalTok{) }\SpecialCharTok{*} \FunctionTok{qnorm}\NormalTok{(}\DecValTok{5} \SpecialCharTok{/} \DecValTok{8}\NormalTok{))) }\SpecialCharTok{*}\NormalTok{ factor}
\NormalTok{  robustbase}\SpecialCharTok{::}\FunctionTok{Qn}\NormalTok{(x, constant)}
\NormalTok{\}}
\end{Highlighting}
\end{Shaded}

\newpage

\printbibliography

\end{document}